\documentclass{iaus}
\usepackage{graphicx}
\usepackage{graphics}

\title[VLT FLAMES integral-field unit observations of PNe] 
{Unravelling the chemical inhomogeneity of PNe with VLT FLAMES integral-field
unit spectroscopy}

\author[Tsamis et al.]   
{Y. G. Tsamis$^1$, J. R. Walsh$^2$, D. P\'equignot$^3$, M. J. Barlow$^1$,
\break X.-W. Liu$^4$, and I. J. Danziger$^5$}

\affiliation{$^1$Dept. of Physics \& Astronomy, University College London,
London WC1E 6BT, U. K.\\
$^2$European Southern Observatory, D-85748 Garching, Germany\\
$^3$LUTH, Observatoire de Meudon-Paris, F-92195 Meudon, France\\
$^4$Dept. of Astronomy, Peking University, Beijing, China\\
$^5$Osservatorio di Trieste, I-34131 Trieste, Italy}

\pubyear{2006}
\volume{234}  
\pagerange{119--126}
\date{?? and in revised form ??}
 \jname{Planetary Nebulae in our Galaxy and Beyond}
\editors{M. J. Barlow \& R. H. M\'endez, eds.}
\begin{document}

\maketitle

\begin{abstract}
Recent weak emission-line long-slit surveys and modelling studies of PNe have
convincingly argued in favour of the existence of an unknown component in the
planetary nebula plasma consisting of cold, hydrogen-deficient gas, as an
explanation for the long-standing recombination-line versus forbidden-line
temperature and abundance discrepancy problems. Here we describe the rationale
and initial results from a detailed spectroscopic study of three Galactic PNe
undertaken with the VLT FLAMES integral-field unit spectrograph, which advances
our knowledge about the small-scale physical properties, chemical abundances
and velocity structure of these objects across a two-dimensional field of view,
and opens up for exploration an uncharted territory in the study and modelling
of PNe and photoionized nebulae in general.

\end{abstract}

\firstsection 
\section{Introduction}

In recent years, extensive weak emission-line surveys of PNe have all focused
on the optical recombination-line (ORL) spectra of carbon, nitrogen, oxygen and
neon ions, and the comparison between abundances derived from ORLs versus those
from the bright collisionally excited lines (CELs) of these objects (Tsamis et
al. 2003b, 2004; Liu et al. 2004; Wesson et al. 2005). These studies show that
for the majority of PNe abundances of the above elements, relative to hydrogen,
from ORLs are a factor of 2--3 higher than the corresponding CEL values. For
about 5--10\% of PNe however, the abundance discrepancy factors (ADF) are in
the range of 4--80. Important correlations have been demonstrated such as those
between the ADFs for a range of ions and the difference between the [O~{\sc
iii}] and Balmer jump temperatures, or between the ADF and the PN radius and
intrinsic brightness, whereby larger/fainter (and arguably older) PNe display
higher ADFs than more compact, young objects. All the above surveys, however,
were based on data secured with spectrographs using long-slits (e.g. ESO 1.52-m
Boller \& Chivens, NTT 3.5-m EMMI) operated either at fixed nebular positions
or in some cases scanned across the nebular surface, thus yielding average
spectra of a given target with the unavoidable loss of spatial information in
the direction perpendicular to the slit. A handful of studies charted the
ORL/CEL abundances, the ADF behaviour and other nebular properties along the
useful slit length for a few PNe (e.g. NGC\,6153, Liu et al. 2000; NGC\,6720,
Garnett \& Dinerstein 2001; NGC\,7009, Krabbe \& Copetti 2006), as well as for
the HII region 30~Doradus (Tsamis et al. 2003a). The spatial resolution of
these however was only moderate (usually worse than 1$''$/pix) and the spectra
did not sample the seeing. The FWHM spectral resolution,
$\lambda$/$\Delta\lambda$ ($=R$), was also not better than $\sim$3000. Similar
studies with \'echelle spectrographs, such as VLT UVES (e.g. Peimbert et al.
2004), fared better in spectral resolution ($\sim$8800), but the spatial
coverage of those was even more restricted.

The advent of integral-field unit (IFU) spectrographs has opened up new options
for PN research (see also Roth, this volume). IFUs provide high spatio-spectral
resolution over a large field of view: with the use of arrays of small
microlenses/fibres the seeing can be sampled and contiguous fibre coverage
prevents flux losses, in contrast to long-slit techniques. Our instrument of
choice, VLT FLAMES in Argus mode, provides coverage of a 12$''\times7''$ or
6.6$''\times4.2''$ field of view at spatial resolutions of 0.52$''$ or
0.30$''$, respectively, and at medium to high $R$. In this work we have used
FLAMES/Argus to observe the Galactic disk PNe NGC 5882, 7009, and 6153. Our
published long-slit studies show that these PNe have ADFs of $\sim$2, 5 and 10,
respectively. It was argued for a variety of reasons, but primarily since the
\emph{ISO} fine-structure CELs yield consistent abundances with the optical
CELs, that small-scale temperature fluctuations do not contribute significantly
to the observed ORL/CEL abundance discrepancies, and the CEL abundances do
provide a representative measure of the gas metallicity in these targets. The
CNONe ORL emission, on the other hand, is best explained by the presence of a
small amount of gas (1--2\% of the total ionized nebular matter) which is
hydrogen-deficient, and therefore cold ($\lesssim$1000\,K), and which does not
emit high-excitation energy UV/optical CELs. Our Argus data allow us to map the
two-dimensional distribution of heavy-element ORL vs. CEL emission across the
PNe in unprecedented detail and are expected to yield new insights into the
astrophysics of these objects.

\section{FLAMES/Argus observations}

\begin{figure}
 \includegraphics[scale=0.45]{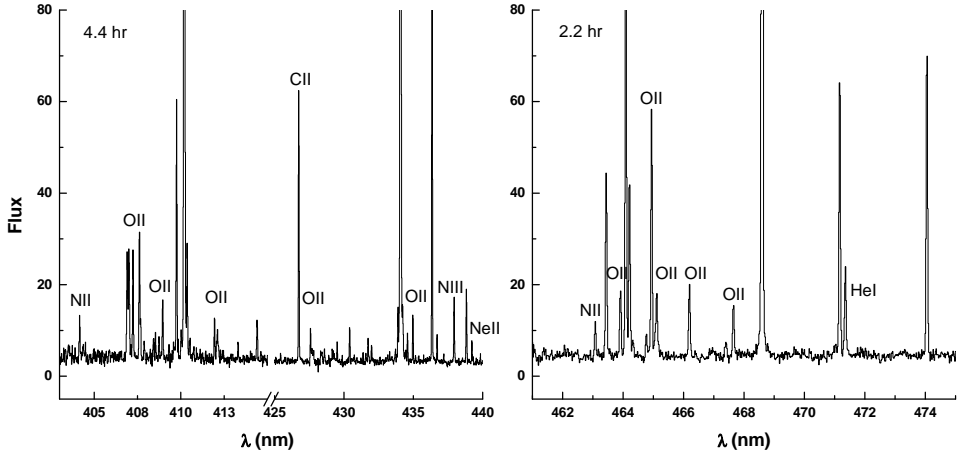}
  \caption{Single 0.52$''$ spaxel FLAMES/Argus spectrum of NGC\,6153 showing the optical
  recombination lines from CNONe ions. Note the full complement of O~{\sc
ii} V1 multiplet ORLs near 4650\,\AA.}\label{fig:wave}
\end{figure}

The spectra were obtained in queue mode between March--August 2005 in subarcsec
seeing. NGC\,7009 was observed with both small and large IFUs, while NGC\,5882
\& NGC\,6153 were observed with the large IFU only. The large IFU covered
roughly one quadrant of the surface of NGC\,6153 \& NGC\,7009, and was placed
along the radial direction so as to sample both inner and outer nebular
regions. The bright inner shell of NGC\,5882 was almost wholly covered with the
12$''\times7''$ IFU (Fig.~2). Spectra of NGC\,7009 (4188--4392,
4538--4759\,\AA) secured with the small IFU have a $R$ of 32,500 ($=$9.2\,km/s)
allowing us to measure gas radial velocities to an accuracy of a few km/s: we
can thus probe whether heavy element ORLs arise from kinematically distinct
regions than CELs in this nebula, by tracking subtle differences in the
velocity profiles of CELs vs. ORLs as a function of position on the
emission-line maps, and via the detection of asymmetrical profiles/faint line
wings (Fig.~3). The 3964--5078\,\AA\ spectra of all 3 PNe taken with the large
IFU have a $R$ of $\sim$10,000--12,000 ($=$25--30\,km/s), comparable to the
typical expansion velocity of a PN and thus optimal for the detection of faint
ORLs. Fig.~1 shows a representative deep spectrum of NGC\,6153 extracted from a
\emph{single} 0.52$''$ spaxel, reaching weak O~{\sc ii} ORLs (of intensity less
than 1\% that of H$\beta$) at a S/N ratio greater than 10.

\section{First results}

\begin{figure}
\centering

\resizebox{4.06cm}{!}{\includegraphics{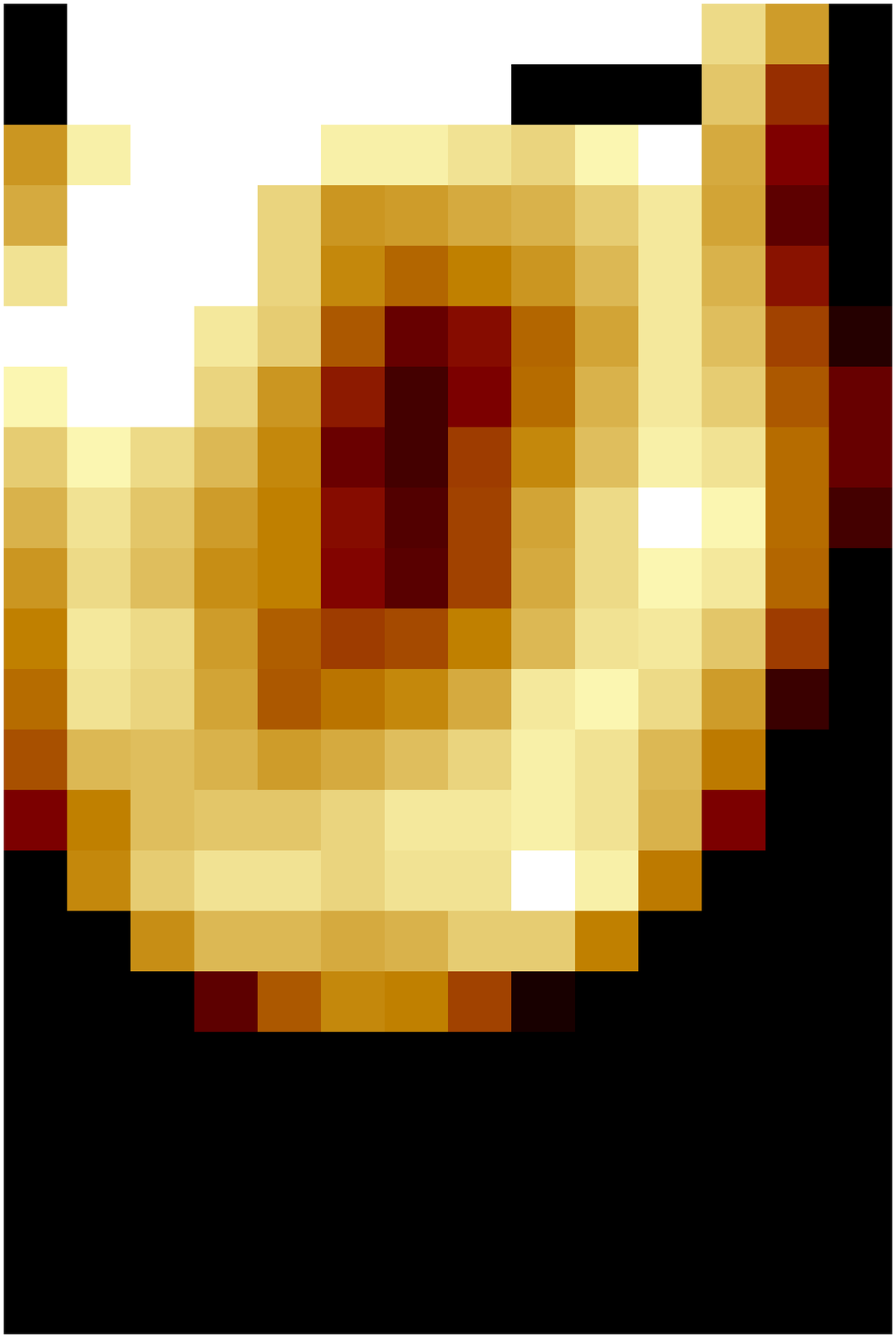} }
\resizebox{4.0cm}{!}{\includegraphics{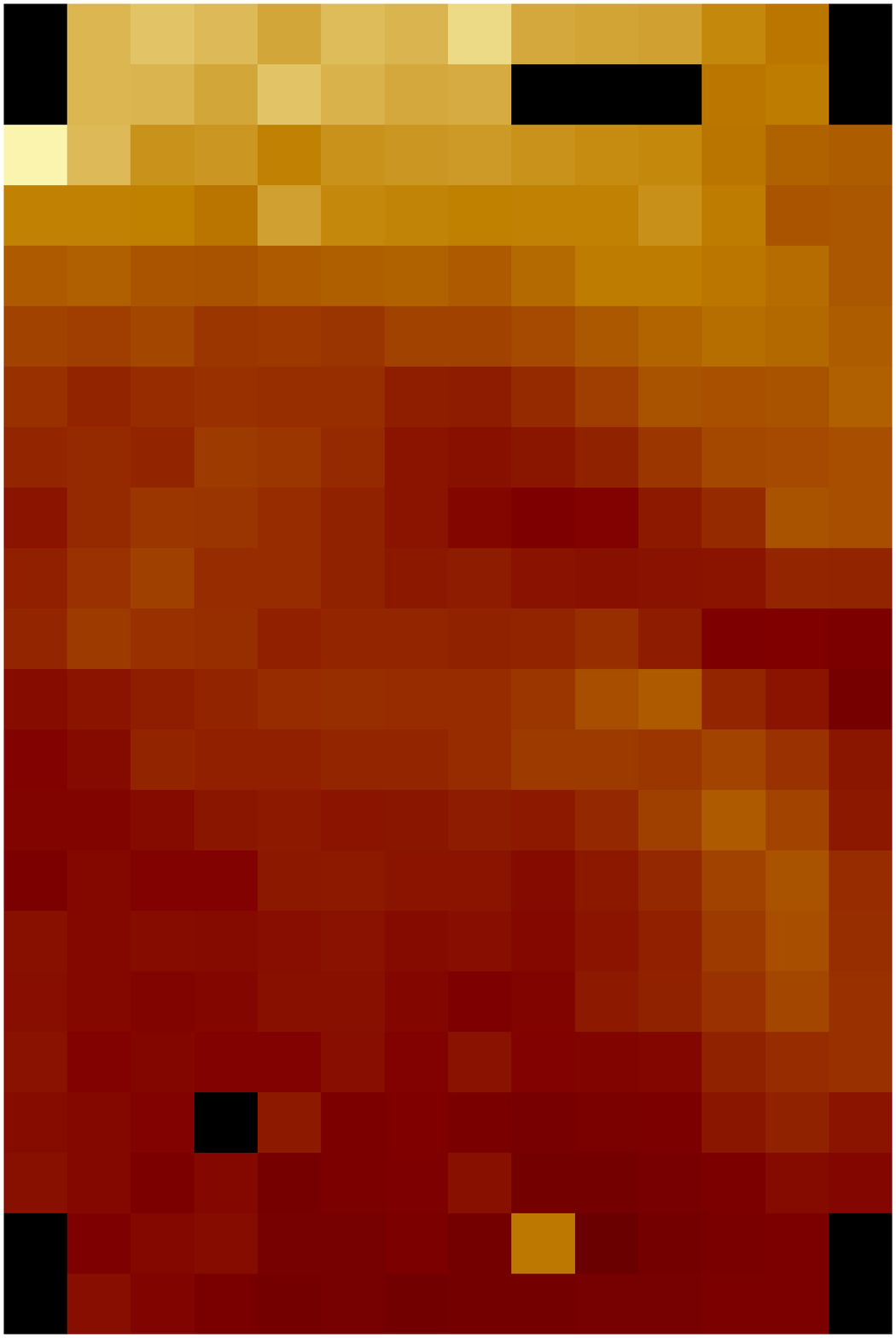} }
\resizebox{4.00cm}{!}{\includegraphics{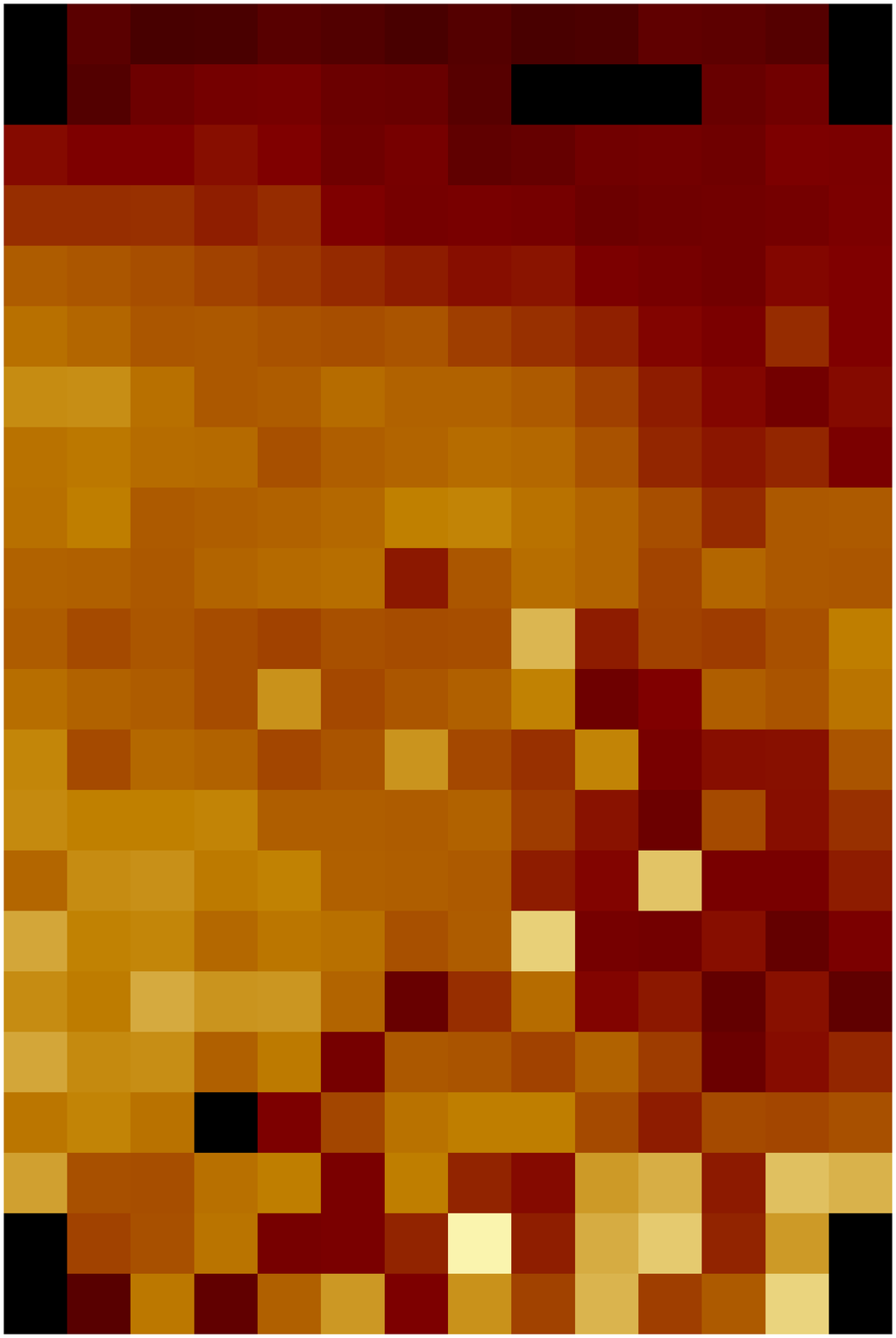} }


\caption[]{(left): NGC\,5882 -- [O~{\sc iii}] $\lambda$4959 map (log-scale);
(centre): NGC\,6153 -- forbidden-line O$^{2+}$/H$^+$ abundance ratio map;
(right): NGC\,6153 -- ORL O$^{2+}$/H$^+$ abundance ratio map. All are
12$''\times7''$ fields. The blank corner pixels correspond to sky-fibres; those
in the 2nd row are dead fibres. See text for details and online version for
colour.}
\end{figure}

The data were reduced with the girBLDRS pipeline provided by the Geneva
Observatory and flux calibrated within IRAF. Custom-made IRAF routines allowed
us to create data cubes and to extract high S/N ratio emission-line maps by
fitting Gaussians to the lines. The maps were then dereddened and converted to
physical quantities, such as electron temperatures ($T_{\rm e}$), densities
($N_{\rm e}$) and abundances. In Fig.~2(\textit{left}) the [O~{\sc iii}]
$\lambda$4959 map of NGC\,5882 is shown. In Fig.~2(\textit{centre}) the
forbidden-line O$^{2+}$/H$^+$ abundance map of the SE quadrant of NGC\,6153 is
shown: it was created using an [O~{\sc iii}] $\lambda$4959/H$\beta$ ratio map,
an [O~{\sc iii}] $\lambda$4363/$\lambda$4959 $T_{\rm e}$ map and an [Ar~{\sc
iv}] $\lambda$4711/$\lambda$4740 $N_{\rm e}$ map. The central star is at the
bottom-right corner. The mean O$^{2+}$/H$^+$ abundance ratio from the map is
3.9$\times$10$^{-4}$, which compares well with the value of
4.3$\times$10$^{-4}$ from the scanning long-slit ESO 1.52-m study of the entire
PN (27$''\times34''$) by Liu et al. (2000). In Fig.~2(\textit{right}) the
recombination-line O$^{2+}$/H$^+$ abundance map of the same region is shown: it
was created from an O~{\sc ii} $\lambda$4649/H$\beta$ ratio map. The mean
O$^{2+}$/H$^+$ value from that map is 4.0$\times$10$^{-3}$ compared to
2.8$\times$10$^{-3}$ from the scanning long-slit study. The derived ADF for
O$^{2+}$ from the ratio of the two NGC\,6153 maps ranges from $\sim$5 to 20
with decreasing nebular radius and peaks around the central star; this directly
implicates the central star itself, or the hard radiation field close to it, in
the ORL/CEL discrepancy problem. Maps such as these will also allow us to track
the pixel to pixel variation of crucial quantities such as, for example, the
$T_{\rm e}$ from the O~{\sc ii} $\lambda4089$/$\lambda4649$ ORL ratio, or the
$N_{\rm e}$ from the intramultiplet intensity ratios of O~{\sc ii} V1 ORLs and
can throw new light on the nature of the hydrogen-poor nebular component.

\begin{figure}
\centering

\includegraphics[scale=0.4, angle=90]{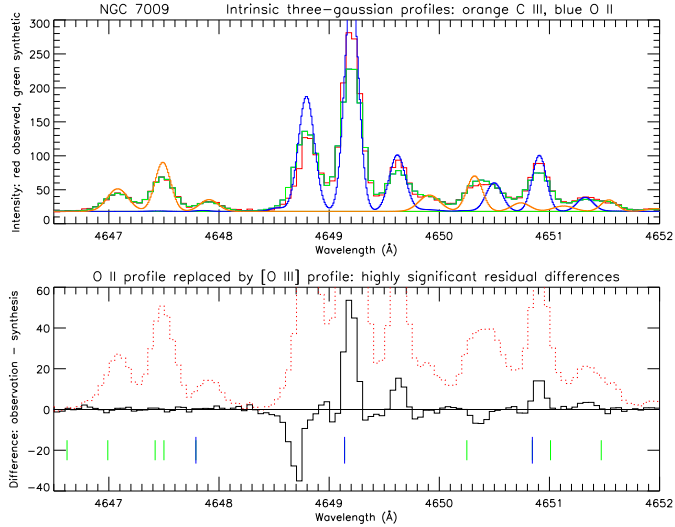}

\caption[]{NGC\,7009: observed and synthetic spectrum of a 0.9$''\times0.9''$
region at $R = $~32,500 extracted from the 6.6$''\times4.2''$ IFU. See text for
details and online version for colour.}

\end{figure}

Fig.~3 shows the high resolution spectrum of a 0.9$''\times0.9''$ region in
NGC\,7009, near the projected edge of the bright inner nebular shell, covering
the C~{\sc iii} 4647.42, 4650.25, 4651.47\,\AA\ and O~{\sc ii} 4649.13,
4650.84\,\AA\ ORLs. Each line exhibits 3 components with central wavelengths at
$\sim-93, -60$ and $-28$\,km/s, probably corresponding to the edge of the inner
PN shell (central peak) and the expanding fainter, outer shell (peaks on either
side). In the upper panel observed (red) and synthetic (green) spectra are
shown. The intrinsic synthetic spectra (before convolution with the
instrumental profile) of C~{\sc iii} (orange) and O~{\sc ii} (blue) are
overplotted. The synthetic O~{\sc ii} line profiles have been replaced by the
profile of [O~{\sc iii}] $\lambda$4363. The bottom panel, which shows the
difference between the observed and synthetic spectra, yields large residuals,
especially for the central O~{\sc ii} peak, i.e. the O~{\sc ii} ORLs have
significantly narrower widths than [O~{\sc iii}] $\lambda$4363. This indicates
that, even though they are emitted from the same O$^{2+}$ ion, the O~{\sc ii}
ORLs and [O~{\sc iii}] $\lambda$4363 cannot originate from material of
identical physical properties.

\end{document}